\documentclass[10pt,twocolumn,conference,letterpaper]{ieeeconf}
\usepackage{amsmath,amssymb,euscript}
\usepackage{latexsym,graphicx,color,amstext,bm}

\IEEEoverridecommandlockouts     

\newtheorem{proposition}{Proposition}[section]
\newtheorem{remark}{Remark}[section]

\newcommand{\ud}{{\rm{d}}}

\newcommand{\I}{{\mathcal I} }
\newcommand{\F}{{\mathcal F} }
\newcommand{\Expect}{{\mathbb{E}}}

\newcommand{\interior}[1]{%
	{\kern0pt#1}^{\mathrm{o}}%
}

\newcommand{\mR}{{\mathbb R}}

\newcommand{\ignore}[1]{}

\newcommand{\W}{{\rm W}}

\newcommand{\black}{\color{black}}

\definecolor{grey}{rgb}{0.6,0.3,0.3}
\definecolor{lgrey}{rgb}{0.9,.7,0.7}

\newcounter{rmnum}

\newcounter{anum}

\title{\huge On the relation between information and power\\ in stochastic thermodynamic engines \thanks{Supported in part by the NSF under grants 1807664, 1839441, 1901599, 1942523, and the AFOSR under FA9550-17-1-0435.}}
\author{  Amirhossein Taghvaei$^{*\dagger}$, Olga Movilla Miangolarra$^{*\dagger}$, Rui Fu$^{*\dagger}$, Yongxin Chen$^{\ddagger}$, and Tryphon T.\ Georgiou$^{\dagger}$
\thanks{$^{\dagger}$Department of Mechanical and Aerospace Engineering, University of California, Irvine, CA; \{rfu2,omovilla,ataghvae,tryphon\}@uci.edu}\\
\thanks{$^{\ddagger}$School of Aerospace Engineering, Georgia Institute of Technology, Atlanta, GA; {yongchen@gatech.edu}}\\
\thanks{$^*$Contributed equally and AT directed the completion of the work.}
}

\begin{document}
	
	\markboth{\today}{}
	
	\maketitle

	\begin{abstract}
		The common saying, that information is power, takes a rigorous form in stochastic thermodynamics, where a quantitative equivalence between the two helps explain the paradox of Maxwell's demon in its ability to reduce entropy. In the present paper, we build on earlier work on the interplay between the relative cost and benefits of information in producing work in cyclic operation of thermodynamic engines (by Sandberg etal. 2014). Specifically, we study the general case of overdamped particles in a time-varying potential (control action) in feedback that utilizes  continuous measurements (nonlinear filtering) of a thermodynamic ensemble, to produce suitable adaptations of the second law of thermodynamics that involve information. 
	\end{abstract}
	
\section{Introduction}

Thermodynamics is the branch of physics which is concerned with the relation between heat and other forms of energy. Historically, it was born of the quest to quantify the maximal efficiency of heat engines, i.e., the maximal ratio of the total work output over the total heat input to a thermodynamic system. This was accomplished in the celebrated work of Carnot ~\cite{carnot1986reflexions,callen1998thermodynamics} where, assuming that transitions take place infinitely slowly, it was shown that the maximal efficiency possible is $\eta_C=1-T_c/T_h$ ({\em Carnot efficiency}), where $T_h$ and $T_c$ are the absolute temperatures of two heat reservoirs, hot and cold respectively, with which the heat engine alternates contact.

 Somewhat inadvertently, Carnot's work gave birth to the second law of thermodynamics, which affirms that the total entropy of a system can never decrease, and whose most prominent consequence is to highlight the arrow of time. Specifically, it states that the work output $-\mathcal W$ can not exceed the free energy difference between the initial and terminal states of the thermodynamic system $-\Delta \F$, that is,
 \begin{equation*}
     \mbox{\fbox{$\mathcal{W}\geq \Delta \F$}}
 \end{equation*}
 In Lord Kelvin's words, the second law of thermodynamics amounts to the impossibility of a self-acting machine, unaided by any external agency, to convey heat from one body to another at a higher temperature \cite{Thomson_1851}.
 
Soon after Lord Kelvin's assertion, Maxwell's far reaching thought experiment that involved a demonic creature \cite{maxwell1871demon}, pointed to ways to generate a temperature gradient by sorting particles in a thermodynamic ensemble based on velocity measurements. The apparent paradox was not resolved until, a century later, Rolf Landauer affirmed that information is physical \cite{Landauer1991ln2}. Starting from the basic assumption that information must be stored somewhere, he was able to link the loss of information with the work performed.


The relation between information and work gradually became a central theme of stochastic thermodynamics~\cite{seifert2008stochastic,sekimoto2010stochastic,seifert2012stochastic,brockett2017thermodynamics,parrondo2015thermodynamics} -- a field shaped in the past two decades to study thermodynamic transitions taking place in finite time.
To this end, thermodynamic ensembles are modeled via stochastic differential equations and notions of work and heat are described at the level of individual trajectories of the ensemble. Ideas form stochastic control were naturally brought in and
%
the second law was extended to include discrete time measurements \cite{sagawa2008discreteqmeas}, as well as continuous ones, both for quantum systems \cite{Mauro2020contqmeas} and classical systems under feedback cooling  \cite{mitter2005information,Horowitz_2014}. In these studies, a generalized version of the second law has taken the form: \begin{equation*}
    \mbox{\fbox{$\mathcal{W}\geq \Delta \F-k_B T \I$}}
\end{equation*} where $\I$ represents the information utilized in effecting a thermodynamic transition. Information engines that work without temperature gradient and only fueled by information soon followed \cite{Abreu_2011workfeedback,Bauer_2012infomachine,parrondo2021extracting}. 

The present work aims to develop further this circle of ideas within a stochastic controls perspective. Specifically, we derive tighter forms of the second law for over-damped systems in general, modeled by Langevin equations and subject to continuous nonlinear measurements. Moreover, in the setting where the ensemble is seen as the medium of a thermodynamic engine and where performance is measured by power drawn, detailed expressions for maximal power and efficiency are derived in the setting of linear-dynamics with Gaussian-distributions.

The exposition proceeds as follows. Section \ref{sec:OMT} provides a preamble on optimal mass transport -- a theory that constitutes the template for optimal control of probabilistic ensembles. 
Section \ref{sec:model} explains the stochastic model of a thermodynamic engine, the
energy exchange mechanism, and the form of the second law in the absence of feedback. 
Section \ref{sec:discrete} extends the second law to the case when information from a single measurement becomes available.
Section \ref{sec:cont} contains our main results on operating a thermodynamic engine with nonlinear continuous time measurements and a form of the second law that applies in this case. Section \ref{sec:V} details expressions for maximal power and efficiency of the linear Gaussian information engine. Finally, Section \ref{sec:CR} provides perspective and research directions.

\section{Preliminaries on optimal mass transport}\label{sec:OMT}
We outline certain geometrical notions from optimal mass transport~\cite{villani2003topics} that play an essential role in the present paper. 
Given probability distributions $p_0$ and $p_f$ on $\mR$, 
\begin{equation}\nonumber
    \W_2(p_0,p_f)^2 := \inf_{\pi \in  \Pi(p_0,p_f)} \int_{\mR\times \mR} |x-y|^2 \pi(x,y) \ud x \ud y ,
\end{equation}
where $\Pi(p_0,p_f)$ denotes the set of joint 
probability distributions on $\mR\times \mR$ with $p_0$, $p_f$ as marginals, defines the so-called $2$-Wasserstein distance (metric).
It turns out that $\W_2(p_0,p_f)$ makes probability distributions into a geodesic space. In turn, geodesics correspond to (optimal) flows between endpoint distributions that provide an alternative expression for $\W_2(p_0,p_f)$. 
Specifically, the time-varying probability distribution $p(t,x)$, driven by the velocity field $v(t,x)$ via the continuity equation $\frac{\partial p}{\partial t}+\nabla \cdot(pv) = 0$. Then
\begin{equation}\label{eq:action}
    \mathcal A[p,v]:= \int_0^{t_f}\int_\mR |v(t,x)|^2 p(t,x) \ud x \ud t,
\end{equation}
represents an action integral for the flow $p(\cdot,x)$. A celebrated result by Benamou and Brenier states 
\begin{align}\label{eq:Benamou}
     &\min_{(p,v) \in \mathcal P(p_0,p_f)}~\mathcal A[p,v] = \frac{1}{t_f}\W_2^2(p_0,p_f),
\end{align}
as a minimal over the set of paths connecting $p_0$ to $p_f$. 


\section{Stochastic thermodynamic model}
\label{sec:model}
In this paper particles are governed by the overdamped  Langevin dynamics (one-dimensional, for simplicity)
%
\begin{equation}\label{eq:Langevin}
\gamma \ud X_t =  - \nabla_x U(t,X_t) \ud t + \sqrt{2\gamma k_BT}\ud B_t\quad X_0 \sim p_0,
\end{equation}  
where $X_t\in \mR$ denotes the location of a particle, $p_0$ the initial distribution of an ensemble, $\gamma$ the viscosity coefficient of the ambient medium, $k_B$ the Boltzmann constant, $T$ the temperature of a heat bath, $B_t$ a standard Brownian motion that models the thermal excitation from the heat bath, and $U(t,x)$ a time-varying potential exerting a force $-\nabla_x U (t,x)$ on a particle at location $x\in \mR$. 
The potential function $U(t,x)$ is  externally controlled and exchanges work with the particle. 
The work performed on the particle, during the interval $[0,t_f]$, is~\cite[Ch. 5]{sekimoto2010stochastic}\footnote{This definition of work is standard in stochastic thermodynamics, but differs from the one in~\cite{sandberg2014maximum}. See also \cite{horowitz2008comment}, \cite{peliti2008comment}, \cite{peliti2008work}, \cite{vilar2008failure}.}
\begin{equation}\label{eq:W-particle}
W = \int_0^{t_f} \partial_t U(t,X_t) \ud t.
\end{equation}
The average work is
\begin{align*}
\mathcal W &= \int_0^{t_f}  \Expect[\partial_t U(t,X_t)] \ud t =  \int_0^{t_f}  \int \partial_t U(t,x) p(t,x) \ud x \ud t,
\end{align*} where the probability $p(t,x)$ of the particle $X_t$ evolves according to the Fokker-Planck equation
\begin{equation*}
\partial_t p = \frac{1}{\gamma}\nabla \cdot(p[\nabla U + k_BT\nabla \log(p)]= -\nabla \cdot(pv),
\end{equation*}
where we introduced the effective velocity field \[v := -\frac{1}{\gamma}(\nabla U + k_B T \nabla \log (p)).\]

In order to state the second law of thermodynamics, we introduce the notion of free energy corresponding to a potential function $U$ and a probability distribution $p$, namely~\cite{parrondo2015thermodynamics},\footnote{This is a notion of non-equilibrium free energy, since $p$ does not need to be the Boltzmann distribution $p\propto \exp({-\frac{U}{k_BT}})$ }
\begin{equation}
\label{eq:free-energy}
\F(U,p) = \int U p\ud x + k_B T \int \log(p) p \ud x.
\end{equation}
The first term represents the energy and the second term represents the negative of entropy, while together, $\F$ relates to the relative entropy between $p$ and the Boltzmann distribution corresponding to the potential. 
The following proposition relates the average work over the interval $[0,t_f]$ to the free energy difference between the initial and final states, giving a version of the second law of thermodynamics. 
\begin{proposition}
    For the over-damped Langevin dynamics~\eqref{eq:Langevin}, the average work satisfies the identity,
   \begin{align}
\label{eq:second-law-eq}
\mathcal W =& \Delta \F+ 
\gamma \int_0^{t_f} \int |v(t,x)|^2 p(t,x) \ud x \ud t,
\end{align} 
    and the bound 
    \begin{equation}
\label{eq:second-law-ineq-W2}
\mbox{\fbox{$\mathcal W \geq \Delta \F + \frac{\gamma}{t_f} \W_2^2(p(0,\cdot),p(t_f,\cdot))$}}
\end{equation} 
where $\Delta \F=\F\left(U(t_f,\cdot),p(t_f,\cdot)\right) -  \F\left(U(0,\cdot),p(0,\cdot)\right)$.
\end{proposition}
\medskip
\begin{remark}
The second term in the identity~\eqref{eq:second-law-eq} is equal to the action integral~\eqref{eq:action} and represents the dissipation along the thermodynamic transition.  According to \eqref{eq:Benamou}, its minimum is the Wasserstein distance between the end-point distributions, concluding~\eqref{eq:second-law-ineq-W2}. The bound is tight and can be achieved by transporting along the geodesic with constant velocity. In the quasi-static limit, as $t_f \to \infty$,  the dissipation term  vanishes, leading to the classical statement of the second law $\mathcal W \geq \Delta \F$. As a result, the bound~\eqref{eq:second-law-ineq-W2} is interpreted as refinement of the second law for finite-time transitions.  It was obtained in~\cite{chen2019stochastic} for Gaussian setting and generalized in~\cite{fu2020maximal} to arbitrary distributions. 
\end{remark}

\section{Single measurement}
\label{sec:discrete}
We now extend the second law (i.e., the bound~\eqref{eq:second-law-ineq-W2}) to the case where access to a single noisy measurement of the particle's location is available. 
Thus, assume we have access to noisy measurement $Y$ of the initial particle location $X_0$. We utilize the measurement $Y$ to modify our control in $U$, denoted $U^Y$. The expected work conditioned on $Y$ is 
\begin{align*}
\mathcal W(Y) =& \int_0^{t_f}  \Expect[\partial_t U^Y(t,X_t)|Y] \ud t. 
\end{align*}
The information in $Y$ allows extracting work, and this additional work is characterized in terms of the mutual information between $X_t$ and $Y$,
\begin{equation}
     \I(X_t;Y) := \mathcal H(X_t) - \mathcal H(X_t|Y).
\end{equation}
Here, $\mathcal H(X_t)$ and $\mathcal H(X_t|Y)$ are the entropy of $X_t$ and the conditional entropy of $X_t$ given $Y$ respectively, defined as
\begin{align*}
   \mathcal  H(X_t) &:= - \int \int \log\left(p_{X_t}(x)\right)p_{X_t}(x) \ud x, \\
    \mathcal H(X_t|Y) &:= - \int \int \log\left(p_{X_t|Y}(x|y)\right)p_{X_t,Y}(x,y) \ud x \ud y,
\end{align*}
where $p_{X_t,Y}$ denotes the joint distribution of $(X_t,Y)$, $p_{X_t|Y}$ the conditional, and $p_{X_t}$ and $p_{Y}$ the marginals. 

The following proposition states an extension of the second law. In order to compare to the case with no measurement, we set the initial and final potential
to a fixed function $U_0$ and $U_f$ respectively. 
Note that  the potential function is allowed to have discontinuous jump at initial and final time.

\begin{proposition}
    Consider a particle governed by the over-damped Langevin dynamics~\eqref{eq:Langevin}, and access to a noisy measurement $Y$ of initial particle location $X_0$. Fix the initial and final potential functions $U_0$ and $U_f$, respectively.  Then, the average work satisfies the bound
    \begin{equation}\label{eq:second-law-Y}
    \mbox{\fbox{$\begin{array}{l}
       \Expect[\mathcal W(Y)]   \geq\Delta \F + \frac{\gamma}{t_f}\W_2^2(p_{X_0},p_{X_{t_f}}) \\[.05in]
         \hspace*{50pt}  - k_BT( \I(X_0;Y)- \I({ X_{t_f}};Y))
    \end{array}$}}
\end{equation}
%
%
where $\Delta \F = \F(U_f,p_{X_{t_f}}) - \F(U_0,p_0)$. 
\end{proposition}
\medskip

\begin{remark}
Compared to~\eqref{eq:second-law-ineq-W2}, the new bound~\eqref{eq:second-law-Y} contains an additional term  $k_BT(\I(X_0;Y)-\I(X_{t_f};Y))$. This term quantifies the amount of information by measuring $Y$ that is actually being used as the particle transitions from $X_0$ to $X_{t_f}$. In the case where the system undergoes cyclic transitions, and therefore $\Delta \F=0$,  the information term  provides the {\em maximum amount of work that can be extracted from a single heat bath with constant temperature using feedback}. The thermodynamic system under such a feedback cycle is referred to as an information machine \cite{Bauer_2012infomachine}.
\end{remark}

\begin{remark}
Compared to the previous bounds in the literature of the form $ \Expect[\mathcal W(Y)] \geq \Delta \F - k_B T \I(X_0;Y)$, e.g. ~\cite[Eq. (1)]{Bauer_2012infomachine}, our bound is tighter and involves two additional terms. The additional term involving the Wasserstein distance characterizes the minimum dissipation in the process. The additional term $k_BT\I(X_f;Y)$ contains the information that has not been used at the end of the process and cannot be transformed to work. Assuming the system converges to a steady state independent of $Y$,  both of these terms will tend to zero as $t_f\to\infty$.
\end{remark}
\medskip

\begin{proof}
The conditional probability distribution $p_{X_t|Y}$ satisfies the the Fokker-Planck equation for $t\geq 0$,
\begin{equation}\label{eq:FPK-Y}
  \partial_t p_{X_t|Y} =  - \nabla \cdot(p_{X_t|Y}v^Y)  
\end{equation}
where 
\begin{align*}
    v^Y(t,x) &= -\frac{1}{\gamma}[\nabla U^Y(t,x) + k_BT\nabla \log(p_{X_t|Y}(x|Y))].
\end{align*}
Upon expressing the derivative of the free energy as 
\begin{align*}
\frac{\ud}{\ud t} \F(U^Y(t,\cdot),p_{X_t|Y}) &=  \int \partial_t U^Y(t,x)p_{X_t|Y}(x|y)\ud x \\&- \gamma \int |v^Y(t,x)|^2 p_{X_t|Y}(x|Y) \ud x,
\end{align*}
and integrating over the time interval $[0,t_f]$,
\begin{align*}
\mathcal W(Y) =& \Delta \F^Y+ \gamma \int_0^{t_f} \int |v^Y(t,x)|^2 p_{X_t|Y}(x|Y) \ud x \ud t,
\end{align*}
where $\Delta \F^Y = \F(U_f,p_{X_{t_f}|Y}) -  \F(U_0,p_{X_0|Y})$.
The expected free energy at the initial time is
\begin{align}\nonumber
    \Expect&[\F(U_0,p_{X_0|Y})] = \int U_0(x)p_{X_0|Y}(x|y) p_Y(y) \ud x \ud y\\\nonumber&+k_BT \int \log(p_{X_0|Y}(x|y)) p_{X_0|Y}(x|y) p_Y(y) \ud x \ud y\\\nonumber&=\int U_0(x)p_{X_0}(x) \ud x - k_B T \mathcal H(X_0|Y) ,
    \\ \label{eq:F-I}
     &= \F(U_0,p_{X_0}) -k_B T \I(X_0;Y)
\end{align}
where we used that $\I(X_0;Y) = \mathcal H(X_0)-\mathcal H(X_0|Y)$. 
Then, with a similar conclusion for the expected free energy at $t_f$,
 \begin{align*}
\Expect[\Delta \F^Y] &= \Delta \F  - k_B T[\I(X_0;Y) - \I({ X_{t_f}};Y)].
\end{align*} 
It now remains to bound the dissipation term from below. For a fixed value of the measurement $Y$,  
\begin{align*}
 \int_0^{t_f} \hspace{-6pt}\int |v^Y(t,x)|^2 p_{X_t|Y}(x|Y) \ud x \ud t \geq \frac{1}{t_f}\W_2^2(p_{X_0|Y},p_{X_{t_f}|Y}),
\end{align*}
because of~\eqref{eq:FPK-Y} and the Benamou-Brenier result~\eqref{eq:Benamou}. 
In addition, the expectation of the Wasserstein distance, over the measurement $Y$, satisfies the lower bound 
\begin{equation*}
\Expect[\W_2^2(p_{X_0|Y},p_{X_{t_f}|Y})] \geq \W_2^2(p_{X_0},p_{X_{t_f}}).
\end{equation*}
This bound is obtained using the standard dual formulation of the Wasserstein distance as a sup over linear functional of the marginals. Interchanging the expectation and sup results in this lower-bound and concludes the result.
\end{proof}

\section{Continuous measurements}
\label{sec:cont}
We now consider the case of having access to a continuous stream of measurement given by 
\begin{equation}\label{eq:obs-model}
\ud Z_t = h(X_t)\ud t + \sigma_v \ud V_t,
\end{equation} 
where $h(\cdot)$ is the observation function, $\{V_t\}$ is a Brownian motion representing the noise in measurements, and $\sigma_v$ is  the strength of noise. We assume that $\{V_t\}$ and $\{B_t\}$ are mutually independent processes.
The expected work conditioned on the measurement history, i.e. the filtration $\mathcal Z_t$ generated by the observation process $\{Z_s;s \in [0,t]\}$, is
\begin{align*}
\mathcal W(\mathcal Z_{t_f}) =& \int_0^{t_f} \Expect[\partial_t U^{\mathcal Z_t}(t,X_t)|\mathcal Z_t]\ud t,
\end{align*}
where we used the notation $U^{\mathcal Z_t}(t,X_t)$ to indicate that the potential function at time $t$ may depend on the history of observations up to that point. Similar to the single measurement case, this  information can be used to extract work from the system.  
The information in the continuous-time setting is characterized by the mutual information between the random processes $X_{0:t_f}$ and $Z_{0:t_f}$. For the particular observation model~\eqref{eq:obs-model},  the mutual information is given by \cite{mutualinf1970TDuncan}
\begin{equation}\label{eq:I-Xt-Zt}
    I(X_{0:t_f};Z_{0:t_f}) = \frac{1}{2\sigma_v^2} \int_0^{t_f} \Expect[|h(X_t) - \hat{h}_t|^2] \ud t,
\end{equation}
where $\hat{h}_t := \Expect[h(X_t)|\mathcal Z_t]$. 
\medskip

\begin{proposition}
    Consider the particle governed by the over-damped Langevin dynamics~\eqref{eq:Langevin} and access to a continuous stream of measurements according to~\eqref{eq:obs-model}. Assume the initial and terminal potential functions are fixed to $U_0$ and $U_f$ respectively.  Then, 
    \begin{equation}\label{eq:second-law-Z}
    \mbox{\fbox{$\begin{array}{l}
       \Expect[\mathcal W(\mathcal Z_{t_f})] \geq \Delta \F +\frac{\gamma}{t_f}\W_2^2(p_{X_0},p_{X_{t_f}})  \\[.05in]
         \hspace*{30pt} - k_B T(\mathcal I(X_{0:t_f};Z_{0:t_f}) - \mathcal I(X_{t_f};Z_{0:t_f} )) 
    \end{array}$}}
    \end{equation}
   %
    where $\Delta \mathcal F=\F(U_f,p_{X_{t_f}}) - \F(U_0,p_{X_0})$.
\end{proposition}
\medskip

\begin{remark}
The notion of information in the continuous measurement case  involves the mutual information between the particle location and the measurement $I(X_{0:t_f};Z_{0:t_f})$, as well as the remaining information  $I(X_{t_f};Z_{0:t_f} )$ that has not been used. This result provides the first and tightest analysis for the role of information for feedback systems under continuous nonlinear observation models.     
\end{remark}
\medskip

\begin{proof}
The conditional probability distribution $p_{X_t|\mathcal Z_t}$ evolves according to the Kushner-Stratonovich equation~\cite{xiong2008introduction}
\begin{equation} \label{eq:K-S}
\ud p_{X_t|\mathcal Z_t}= -\nabla \cdot(p_{X_t|\mathcal Z_t} v^{\mathcal Z_t}) \ud t + \frac{1}{\sigma_v^2}p_{X_t|\mathcal Z_t}(h-\hat{h})\ud \xi_t.
\end{equation}  
where $\ud \xi_t=\ud Z_t-\hat{h}_t\ud t$ is the innovation process and
\[v^{\mathcal Z_t} = -\frac{1}{\gamma}[\nabla U^{\mathcal Z_t} + k_BT\nabla \log(p_{X_t|\mathcal Z_t})].\]
Differentiating the free energy 
\begin{align*}
\ud &\F(U^{\mathcal Z_t}(t,\cdot),p_{X_t|\mathcal Z_t}) 
= \bigg[\int \partial_t U^{\mathcal Z_t} p_{X_t|\mathcal Z_t} \ud x \\&- \gamma \int |v^{\mathcal Z_t}|^2p_{X_t|\mathcal Z_t} \ud x + \frac{k_B T}{2\sigma_v^2} \int (h-\hat{h}_t)^2 p_{X_t|\mathcal Z_t} \ud x \bigg]\ud t \\&+
\frac{1}{\sigma_v^2}\left[\int  (U^{\mathcal Z_t} + k_B T \log(p_{X_t|\mathcal Z_t} ))   p_{X_t|\mathcal Z_t} (h-\hat{h}) \ud x \right]\ud \xi_t.
 \end{align*}
  Integrating over the interval and taking the expectation yields 
 \begin{align*}
    &\Expect[\mathcal W(\mathcal Z_{t_f})]  = \Expect[\Delta \F^{\mathcal Z}]- \frac{k_B T}{2\sigma_v^2} \int \Expect[(h(X_t)-\hat{h}_t)^2|] \ud t \\~&  + \gamma \int_0^{t_f} \Expect[|v^{\mathcal Z_t}(t,X_t)|^2]\ud t,
 \end{align*}
 where $\Delta\F^{\mathcal Z} = \F(U_f,p_{X_{t_f}|\mathcal Z_{t_f}}) - \F(U_0,p_{X_0})$ and we used the fact that $ \xi_t $ behaves as a Brownian motion under conditional expectation~\cite[Lemma 5.6]{xiong2008introduction}. 
Using the definition~\eqref{eq:I-Xt-Zt} and applying the relationship~\eqref{eq:F-I} for the expected free energy at the final time concludes
  \begin{align*}
    \Expect[\mathcal W(\mathcal Z_{t_f})]  = &\Delta \F - k_B T(\I(X_{0:t_f};Z_{0:t_f})-\I(X_{t_f};Z_{0:t_f}))\\& + \gamma \int_0^{t_f} \Expect[|v^{\mathcal Z_t}(t,X_t)|^2]\ud t.
 \end{align*}

 It remains to obtain a lower-bound on the dissipation term. By Jensen's inequality  
 \begin{align*}
     \Expect[|v^{\mathcal Z_t}(t,X_t)|^2|X_t] &\geq | \Expect[v^{\mathcal Z_t}(t,X_t)|X_t] |^2
     =| \bar{v}(t,X_t)|^2,
 \end{align*}
where we introduced $\bar{v}(t,x) := \Expect[v^{\mathcal Z_t}(t,X_t)|X_t=x] $. Upon taking the expectation and integrating over the time interval, 
  \begin{align*}
     \int_0^{t_f}\Expect[|v^{\mathcal Z_t}(t,X_t)|^2]\ud t &\geq \int_0^{t_f}\Expect[| \bar{v}(t,X_t)|^2] \ud t.
 \end{align*}
 The proof  follows by showing that  the velocity field $\bar{v}(t,x)$ generates the flow for the marginal distribution $p_{X_t}$, i.e.\ that $ \partial_t p_{X_t} = - \nabla \cdot(p_{X_t} \bar{v})$, to conclude 
   \begin{align*}
     \int_0^{t_f}\Expect[| \bar{v}(t,X_t)|^2]&\geq \frac{1}{t_f}\W_2^2(p_{X_0},p_{X_{t_f}}).
 \end{align*}
 In order to do so, we take the  expectation of both sides of equation~\eqref{eq:K-S} and use the identities 
\begin{align*}
    p_{X_t}(x) &= \Expect[p_{X_t|\mathcal Z_t}(x|Z_{0:t})] \\
    p_{X_t}(x) \bar{v}(t,x) &= \Expect[p_{X_t|\mathcal Z_t}(x|Z_{0:t}) v^{\mathcal Z_t}(t,x)]
\end{align*}
as well as cancel the mean-zero term multiplied by $\ud \xi_t$. 
\end{proof}

\subsection{Efficiency for information engines}

The efficiency for information engines is defined~\cite{Bauer_2012infomachine} as
the ratio between the work output and the amount of information that is available to be used. Thus, in our case,
\begin{equation}
    \eta:=\frac{-\Expect[\mathcal W(\mathcal Z_{t_f})] 
    }{k_BT  \I(X_{0:t_f}; Z_{0:t_f})}.
\end{equation}
In light of~\eqref{eq:second-law-Z}, the efficiency is always smaller than $1$. 
It is also noted that, in order to achieve maximal efficiency, 
it is necessary that $ \I(X_{t_f}; Z_{0:t_f})=0$, and thereby, that all available information has been used within the interval.
 \black
 
\section{Linear Gaussian setting} 
\label{sec:V}
We now focus on the case of a quadratic potential function $U(t,x) = \frac{q_0}{2}(x-r_t)^2$, where the location  $r_t$ of the center of the potential represents the control input while the intensity $q_0$ remains constant. We assume access to continuous measurements of the particle with observation function $h(x)=x$.  Thus, the dynamics for the particle and the observation are
 \begin{subequations}\label{eq:linear-Gaussian}
 	\begin{align}\label{eq:dynamics-linear-Gaussian}
 	\ud X_t &= -\frac{q_0}{\gamma}(X_t-r_t)\ud t + \sqrt{\frac{2k_BT}{\gamma}}\ud B_t\\\label{eq:observation-linear-Gaussian}
 	\ud Z_t &= X_t \ud t + \sigma_v \ud V_t.
 	\end{align}
 \end{subequations}
 
The objective is to maximize the work output during a cycle of period $t_f$ by designing the control input $r_t$.
We assume boundary condition $r_0=r_{t_f}=0$. We also assume that the initial probability distribution is at equilibrium to disregard any amount of work that can be extracted if the system is not prepared at equilibrium. For the initial potential $U_0(x)  = \frac{q_0}{2}x^2$, the equilibrium distribution is Gaussian $N(0,\Sigma_0)$ with variance $\Sigma_0 = \frac{k_BT}{q_0}$. 

In this special linear Gaussian case, the conditional probability distribution of $X_t$ given the observations is Gaussian $N(m_t,\Sigma_t)$, where the mean and variance evolve according to Kalman-Bucy filter equations~\cite{kalman-bucy}
 \begin{subequations}\label{eq:KF}
	\begin{align}
	\ud m_t &= -\frac{q_0}{\gamma}(m_t-r_t)\ud t + \frac{\Sigma_t}{\sigma_v^2}\ud \xi_t \\
	\dot{\Sigma}_t &= -\frac{2q_0}{\gamma}\Sigma_t + \frac{2k_BT}{\gamma} - \frac{1}{\sigma_v^2}\Sigma_t^2,
	\end{align}
\end{subequations}
and $\ud \xi_t = \ud Z_t-m_t\ud t$ is the innovation process. 
 In this special case, the work input to the system is
 \begin{equation*}
 W = \int_0^{t_f}q_0(r_t-X_t)\dot{r}_t \ud t,
 \end{equation*}
 and the conditional expectation of work given the observations is  
  \begin{equation*}
 \mathcal W(\mathcal Z) = \int_0^{t_f}q_0(r_t-m_t)\dot{r}_t \ud t,
 \end{equation*}
 where we replaced $X_t$ with its conditional expectation $m_t$.
 Upon integration by parts and utilizing the boundary conditions $r_0=r_{t_f}=0$,
   \begin{align*}
 \mathcal W(\mathcal Z) 
 =& - \frac{q_0^2}{\gamma}\int_0^{t_f} r_t(m_t-r_t) \ud t + \frac{q_0}{\sigma_w^2} \int_0^{t_f} r_t\Sigma_t \ud \xi_t.
 \end{align*}
Finally, taking expectation, the second term disappears and, in order to maximize work output,
we end up with the following stochastic optimal control problem 
\begin{subequations}
	\begin{align}
	&\min_{u} \frac{q_0^2}{\gamma}\Expect\left[\int_0^{t_f} (u_t^2 - \frac{1}{4}m_t^2) \ud t \right] \\
	\text{s.t.}\quad 	&\ud m_t = -\frac{q_0}{2\gamma}m_t\ud t +\frac{q_0}{\gamma}u_t \ud t+ \frac{\Sigma_t}{\sigma_v^2}\ud \xi_t.
	\end{align}
\end{subequations}
where we introduced the control input $u_t = r_t - \frac{1}{2}m_t$.  
The solution to the stochastic optimal control problem is presented in the following proposition. 
\medskip

\begin{proposition}
	Consider a particle governed by the over-damped Langevin equation with quadratic potential and a linear observation model~\eqref{eq:linear-Gaussian}, and assume that the boundary conditions $r_0=r_{t_f}=0$ and the equilibrium initial distribution $N(0,\frac{k_BT}{q_0})$ hold.  The maximum work output over  $[0,t_f]$ is
	\begin{equation}
	-\mathcal{W}^* = -\frac{q_0}{\sigma_v^2} \int_{0}^{t_f}P_t\Sigma_t^2 \ud t,
	\end{equation}
	and the optimal control is given by $r_t=(\frac{1}{2}-{P}_t)m_t$, where $m_t$ and $\Sigma_t$ are the conditional mean and variance of $X_t$ given by the Kalman-Bucy filter equations~\eqref{eq:KF} and 
	\begin{equation}
	\frac{\gamma}{q_0}\dot{P}_t = ( {P}_t+\frac{1}{2} )^2,\quad {P}_{t_f}=0,
	\end{equation}  
	or, in closed form, $P_t=\big[\frac{q_0(t_f-t)}{\gamma}+2\big]^{-1}-\frac{1}{2}$. Moreover, the efficiency at maximum power is
\begin{equation*}
    \eta= \frac{ -2q_0\int_{0}^{t_f}\bar{P}_t\Sigma_t^2 \ud t}{k_BT\int_0^{t_f} \Sigma_t \ud t}.
\end{equation*}
\end{proposition}
\medskip

\begin{proof}
We use the following candidate value function $\mathcal V(t,m) = P_tm^2 + Q_t$ where $P_t$ and $Q_t$ are time varying parameters to be determined later. Express the objective function as $ \frac{q_0^2}{\gamma}\Expect[J]$ where $J:=\int_0^{t_f} (u_t^2 - \frac{1}{4}m_t^2) \ud t$.  Upon adding the zero term $ \int_{0}^{t_f} \ud \mathcal V(t,m_t)  - \mathcal V(t_f,m_t) + \mathcal V(0,m_0) =0$ to $J$,
and  using 
\begin{align*}
\ud \mathcal V(t,m_t) 
=& \dot{P}_tm_t^2\ud t + \dot{Q}_t \ud t - \frac{q_0}{\gamma}P_t m_t^2 \ud t \\&+ 2\frac{q_0}{\gamma}P_tm_tu_t \ud t+  P_t\frac{\Sigma_t^2}{\sigma_v^2}\ud t + 2P_tm_t \frac{\Sigma_t}{\sigma_v^2}\ud \xi_t,
\end{align*}
we arrive at 
\begin{align*}
J =& \int_0^{t_f} \left[ u_t^2 + 2\frac{q_0}{\gamma}P_tm_tu_t +  (-\frac{1}{4} + \dot{P}_t - \frac{q_0}{\gamma}P_t)m_t^2\right]\ud t \\&+  \int_0^{t_f} (\dot{Q}_t +  P_t\frac{\Sigma_t^2}{\sigma_v^2} )\ud t \\&+\int_0^{t_f}  2P_tm_t \frac{\Sigma_t}{\sigma_v^2}\ud \xi_t- \mathcal V(t_f,m_t) +\mathcal V(0,m_0).
\end{align*}
Now we use our freedom to specify $P_t$ and $Q_t$ to make the first term a complete square and second term zero.
\begin{align*}
\dot{P}_t &= \frac{1}{4} + \frac{q_0}{\gamma} P_t + \frac{q_0^2}{\gamma^2}P_t^2,\quad P_{t_f}=0\\
\dot{Q}_t &= - P_t\frac{\Sigma_t^2}{\sigma_v^2},\quad Q_{t_f}=0.
\end{align*}
We also set the terminal condition to zero to make $\mathcal V(t_f,m_{t_f})=0$. 
The resulting expression for $J$, after taking the expectation,  is 
\begin{align*}
\Expect[J] &= \Expect\left[\int_0^{t_f} \left( u_t+\frac{q_0}{\gamma}P_tm_t\right)^2\ud t + \mathcal V(0,m_0) \right].
\end{align*}
The term $\mathcal V(0,m_0)$ does not depend on $u$. Therefore, the optimal control is $u_t = -\frac{q_0}{\gamma}P_tm_t$, and the optimal value is
\begin{align*}
\mathcal W^*=  \frac{q_0^2}{\gamma}\Expect[J] &= \frac{q_0^2}{\gamma} \mathcal V(0,m_0) =  \frac{q_0^2}{\gamma} Q_0,
\end{align*}
where we used $m_0=0$. The result of the proposition follows by noting $Q_0 = \int_0^{t_f} P_t\frac{\Sigma_t^2}{\sigma_v^2} \ud t$ and changing $P_t \to \frac{q_0}{\gamma}P_t$.
\end{proof}

\begin{figure}[t]
\centering
\includegraphics[width=0.44\textwidth,trim= 5pt 20pt 5pt 10pt]{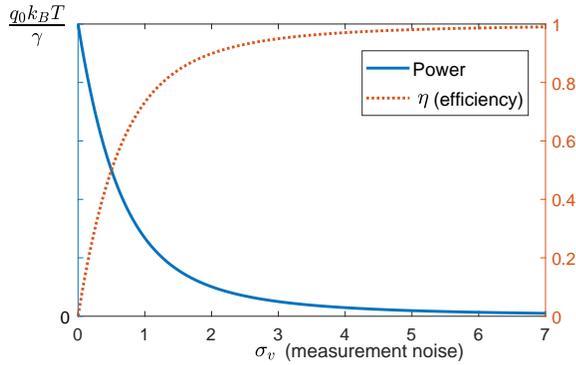}
\caption{Steady-state values for maximum power and efficiency for a linear Gaussian over-damped information machine, as a function of the measurement noise.}
\label{fig}
\end{figure}

\medskip 
\begin{remark}[Steady-state analysis]
Explicit formulas for average power $\frac{-\mathcal W^*}{t_f}$ and efficiency is obtained in steady-state as $t_f \to \infty$. The steady-state average power is
\begin{align*}
\lim_{t_f \to \infty }\frac{-\mathcal W^*}{t_f} &=  
    \frac{q_0k_BT}{\gamma} \frac{1}{\text{SNR}}\bigg( \sqrt{1 + \text{SNR}} -1\bigg)^2.
\end{align*}
where $\text{SNR}= \frac{2\gamma k_B T}{q_0^2\sigma_v^2}$ represents the signal to noise ratio. The limit is obtained using the  steady-state values $P_{ss}=-\frac{1}{2}$ and $\Sigma_{ss}=\sigma_v^2(-\frac{q_0}{\gamma} + \sqrt{\frac{q_0^2}{\gamma^2} + \frac{2k_BT}{\gamma \sigma_v^2}})$. 
In particular, as $\sigma_v \to \infty$,  power converges to zero (because in this case, effectively, no information is available), and as $\sigma_v\to0$, power attains its maximum value $\frac{q_0k_BT}{\gamma}$.  
The efficiency at steady state becomes
\begin{align*}
\lim_{t_f \to \infty}\eta = 
\frac{2}{\text{SNR}}\bigg( \sqrt{1 + \text{SNR}} -1 \bigg)
\end{align*}
Note that as $\sigma_v \to \infty$, the efficiency goes to $1$. However, as $\sigma_v \to 0$, the efficiency converges to $0$, since the available information is infinite. We numerically illustrate power and efficiency tradeoffs as functions of $\sigma_v$ in Figure \ref{fig}.
\end{remark}

\begin{remark} The work presented in this section parallels the work of Sandberg etal.\ \cite{sandberg2014maximum}, but the model and approach are fundamentally different. A major difference is on definition of work \eqref{eq:W-particle} as well as the nature of the control variable.
In spite of the differences, we arrive at the qualitatively similar results on power and  efficiency  (c.f. \cite[Figure 3]{sandberg2014maximum}).
\end{remark}

	\section{Concluding remarks}\label{sec:CR}
Following Rolf Landauer's insight, that information is physical \cite{Landauer1991ln2}, it is no surprise that it can be traded for work. From this vantage point several authors sought to quantify the relation between work, heat, dissipation and information (e.g., \cite{sagawa2010generalized,sandberg2014maximum,Abreu_2011workfeedback,Bauer_2012infomachine,parrondo2015thermodynamics}). The present work follows a similar endeavor.
To this end, we obtained bounds on the maximal amount of work that can be drawn from a thermodynamic ensemble that is in contact with a heat bath of fixed temperature and where information becomes available at one point in time, or when the ensemble is continuously being monitored over a finite interval. Our development brought in new tools and concepts from optimal mass transport and nonlinear filtering. It is hoped that this framework would allow insights on how to achieve tight bounds and derive the corresponding optimal control laws in the general setting, beyond the linear-Gaussian case. It is also of interest to treat under-damped Langevin dynamics and the general case where the temperature of the heat bath varies over time.
\bibliographystyle{IEEEtran}
\bibliography{refs}

\end{document}